# HARMONI: the ELT's First-Light Near-infrared and Visible Integral Field Spectrograph

Niranjan Thatte[1]
Matthias Tecza[1]
Hermine Schnetler[2]
Benoit Neichel[3]
Dave Melotte[2]
Thierry Fusco[3,4]
Vanessa Ferraro-Wood[1]
Fraser Clarke[1]
Ian Bryson[2]
Kieran O'Brien[5]
Mario Mateo[6]
Begoña Garcia Lorenzo[7]
Chris Evans[2]
Nicolas Bouché[8]
Santiago Arribas[9]
and the HARMONI Consortium[a]

[1] Department of Physics, University of Oxford, UK
[2] United Kingdom Astronomy Technology Centre (UKATC), Edinburgh, UK
[3] Laboratoire d'Astrophysique de Marseille (LAM), France
[4] Département d'Optique et Techniques Avancées (DOTA), Office National d'Etudes et de Recherches Aérospatial (ONERA), Paris, France
[5] Physics Department, Durham University, UK
[6] Department of Astronomy, University of Michigan, USA
[7] Instituto de Astrofísica de Canarias (IAC) and Departamento de Astrofísica, Universidad de La Laguna, Tenerife, Spain
[8] Centre de Recherche Astrophysique de Lyon (CRAL), France
[9] Centro de Astrobiología – Instituto Nacional de Técnica Aeroespacial, Consejo Superior de Investigaciones Científicas (CAB-INTA/CSIC), Madrid, Spain

**The High Angular Resolution Monolithic Optical and Near-infrared Integral field spectrograph (HARMONI) is the visible and near-infrared (NIR), adaptive-optics-assisted, integral field spectrograph for ESO's Extremely Large Telescope (ELT). It will have both a single-conjugate adaptive optics (SCAO) mode (using a single bright natural guide star) and a laser tomographic adaptive optics (LTAO) mode (using multiple laser guide stars), providing near diffraction-limited hyper-spectral imaging. A unique high-contrast adaptive optics with high performance and good sky coverage, respectively (AO) capability has recently been added for exoplanet characterisation. A large detector complement of eight HAWAII-4RG arrays, four choices of spaxel scale, and 11 grating choices with resolving powers ranging from *R* ~ 3000 to *R* ~ 17 000 make HARMONI a very versatile instrument that can cater to a wide range of observing programmes.**

## About HARMONI

HARMONI will provide the ELT's workhorse spectroscopic capability at first light. A visible and near-infrared integral field spectrograph (IFS), it provides a "point-and-shoot" capability to simultaneously obtain a spectrum of every spaxel[b] over a modest field of view. Several different flavours of adaptive optics ensure (near) diffraction-limited spatial resolution of ~ 10 milliarcseconds over most of the sky. ELT+HARMONI will transform the landscape of observational astronomy by providing a big leap in sensitivity and resolution — a combination of the ELT's huge collecting area, the exquisite spatial resolution provided by the AO, and large instantaneous wavelength coverage coupled with a range of spectral resolving powers (*R* ~ 3000 to 17 000).

Over the last couple of years, HARMONI has added substantially to the core instrument. The LTAO capability is part of the baseline, as is a high-contrast AO (HCAO) mode that aims to enable direct spectroscopy of extra-solar planetary companions. The University of Michigan has joined as a new partner, providing a much needed cash injection, while the Institut de Planétologie et d'Astrophysique de Grenoble (IPAG) is funding the hardware for HCAO.

HARMONI is equally suited to spatially resolved spectroscopy of extended targets and of point sources, particularly if their positions are not precisely known (for example, transients), or if they are located in crowded fields. The data cube obtained from a single integral field exposure can yield information about the source morphology (via broad- or narrowband images), spatially resolved kinematics and dynamics (via Doppler shifts and widths), chemical abundances and composition (via emission and absorption line ratios) and the physical conditions (temperature, density, presence of shocks) of the emitting region (via line diagnostics). In addition, specialist capabilities such as molecular mapping for high contrast observations, or the use of deconvolution with knowledge of the point spread function (PSF) from AO telemetry extend the areas where HARMONI will make a huge impact. Some examples are showcased in the last section of this article.

## Spatial and spectral grasp

Figure 1a shows the spatial layout of the HARMONI field of view (FoV) at its four different spaxel scales, one of which may be selected *on the fly*. At any spaxel scale, HARMONI simultaneously observes spectra of ~ 31 000 spaxels in a contiguous rectangular field. The common wavelength range in each data cube is ~ 3700 pixels long, after accounting for the stagger between adjacent slitlets and slit curvature. The spaxel scales range from 0.06 × 0.03 arcseconds per spaxel, limited by the focal ratios achievable in the spectrograph cameras, to 4 × 4 milliarcseconds per spaxel, set to Nyquist sample the ELT's diffraction limit in the NIR *H* band. Two other intermediate scales of 10 × 10 milliarcseconds per spaxel and 20 × 20 milliarcseconds per spaxel allow the user to optimise for sensitivity, spatial resolution or FoV, as required. A larger FoV is particularly desirable when using the "nod-on-IFU" technique to achieve accurate sky background subtraction, as it involves positioning the object alternately in each half of the FoV.

The versatility in choice of plate scale is complemented by a large choice of wavelength ranges and spectral resolving powers, as shown in Figure 1b. HARMONI uses Volume Phase Holographic (VPH) gratings for high efficiency. Each grating has a fixed wavelength range, so needs to be physically exchanged to change observing band. One of eleven different gratings can be chosen, which between them provide three different resolving powers (*R* ~ 3000, 7000 and 17 000) spanning the various atmospheric windows in the NIR (atmospheric transmission is shown in grey in Figure 1b).

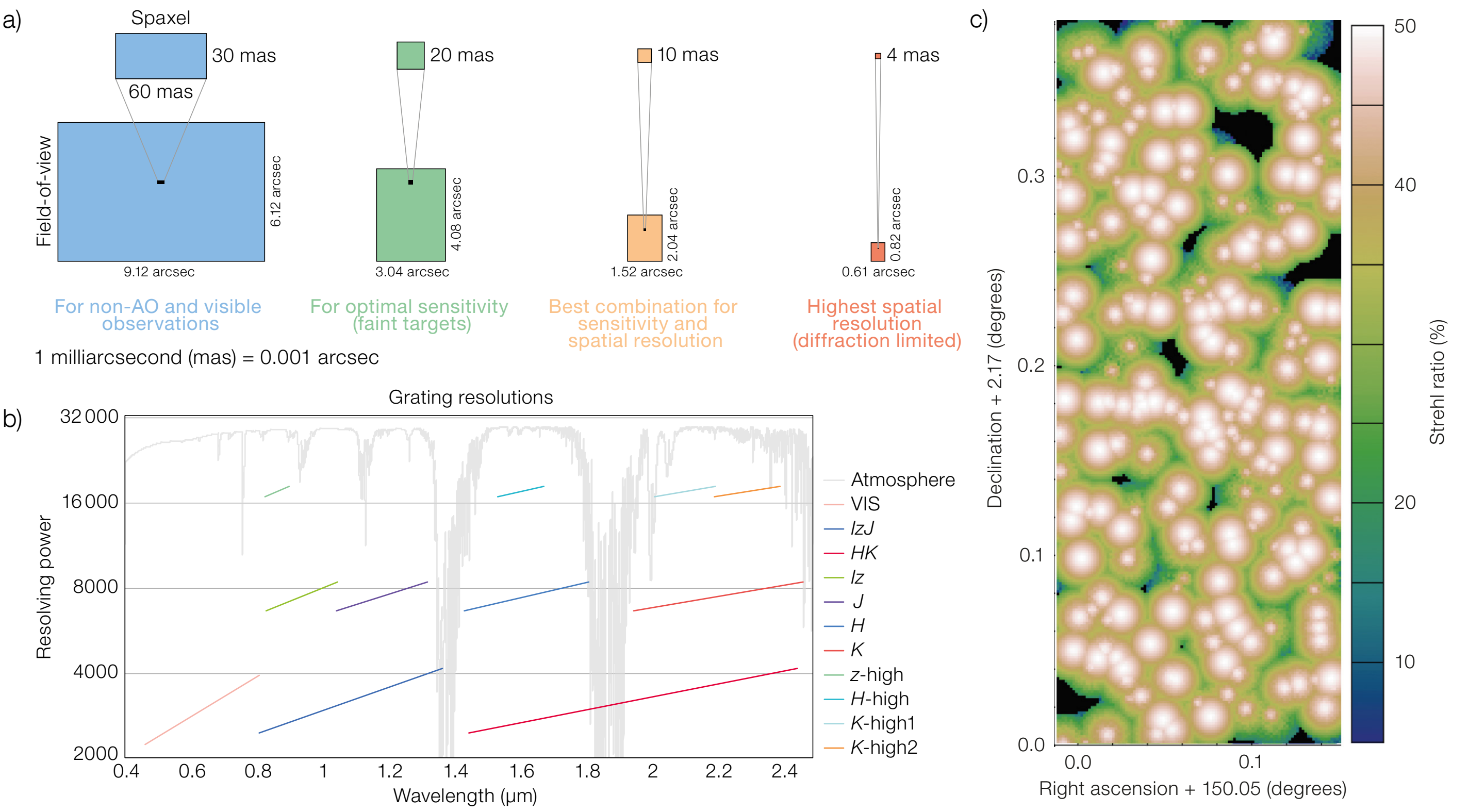

Figure 1. a) Spatial layout of the HARMONI science field, showing the spaxel sizes and fields of view at the four different spaxel scales. b) Spectral coverage and resolving power ranges for each of the 11 HARMONI grating choices. The atmospheric transmission is shown in grey. c) Expected AO performance (Strehl ratio) for the COSMOS deep field, observed with HARMONI LTAO in good seeing conditions (0.43 arcseconds), illustrating the sky coverage achieved for a typical patch of sky.

A fixed-length spectrum implies a natural compromise between instantaneous wavelength coverage and resolving power. One grating provides coverage at visible wavelengths (*V* and *R* bands), requiring a different set of detectors (CCDs instead of the HgCdTe arrays used in the NIR). However, as AO correction works well only at longer NIR wavelengths, the spatial resolution achieved at visible wavelengths is close to seeing-limited, making the large spaxel count somewhat superfluous. Consequently, only half the FoV is offered at visible wavelengths, at all spaxel scales.

## Adaptive optics flavours

The ELT is an adaptive telescope, with M4 (a deformable mirror with over 5000 actuators) and M5 (a fast tip-tilt mirror) providing active correction of atmospheric turbulence. The sensing of the wavefront aberrations is done by the science instruments — better rejection of common-mode disturbances such as flexure and vibrations is achieved by splitting the wavefront sensing light as close to the science focal plane as possible. The scheme used for wavefront sensing leads to HARMONI's four distinct operating modes: LTAO, SCAO, HCAO, and noAO — the last providing no adaptive optics correction of atmospheric turbulence.

In LTAO operation, six laser guide star (LGS) sensors, each with 78 × 78 subapertures, measure the wavefront aberrations at 500 Hz from six sodium laser stars. The laser stars are located in an asterism with a diameter of ~ 1 arcminute, which provides the best compromise between peak performance and robustness to changing atmospheric parameters. HARMONI's AO Control System (AOCS) stitches together the wavefront information from the six lines of sight to reconstruct the wavefront aberration for the on-axis path, and commands M4 and M5 to the appropriate shapes to eliminate the effect of the turbulence, providing a near diffraction-limited corrected wavefront to the IFS.

It is not possible to measure the image motion with LGS, so a separate natural guide star (NGS) is needed to sense tip-tilt and focus. A single off-axis NGS is sensed by HARMONI's NGS System (NGSS), with a probe arm that patrols a 1-arcminute-radius field centred on the IFS FoV. The NGS position and focus are sensed at several hundred Hz in the *H* and *K* bands, while a slow "Truth Sensor" uses the *J*-band light from the same star to eliminate any low-order wavefront errors introduced by the LGS. The NGSS is able to operate with stars as faint as $H_{AB}$ = 19, so that HARMONI's LTAO system can provide excellent sky coverage — 75% of the sky at the south Galactic pole (SGP) with Strehl exceeding 30% in the *K* band under median conditions of atmospheric turbulence (see Figure 1c for an example of LTAO sky coverage).

Even better performance may be obtained by using HARMONI's SCAO system, provided a single, bright, natural guide star is present within 15 arcseconds of the science target of interest. SCAO can also deal with extended objects as AO reference "stars", with slightly degraded performance, as long as the reference is less than 2.5 arcseconds in diameter. Unlike the LTAO system (which uses an off-axis NGS), SCAO uses a dichroic that sends light in the 700–1000 nm range to a pyramid wavefront sensor operating at 500 Hz, with longer wavelengths (1000–2450 nm) available for spectroscopy with the IFS. Both on-axis and off-axis NGS may be used. Optimal performance is achieved for stars down to $V = 12$, with a limiting magnitude of $V \sim 17$. A second SCAO dichroic is available, albeit with a reduced patrol field of 4 arcseconds in diameter, with a cut-in wavelength of 800 nm for spectroscopy, allowing observations that use *z*-band stellar absorption features as diagnostics.

The HCAO mode adds a high-contrast capability to HARMONI, using a combination of a pupil-plane apodiser and a focal-plane mask. Because of uncorrected atmospheric differential refraction (chromatic beam shift), it is not possible to use classical coronagraphs to improve contrast. The novel design by Carlotti et al. (2018) achieves good rejection of starlight — the goal being (post-processed) contrasts of > $10^6$ at separations < 0.2 arcseconds — whilst enabling inner working angles (IWA) of less than 100 milliarcseconds for IFS spectroscopy. HCAO works only with an on-axis NGS. It uses the pyramid wavefront sensor of the SCAO system for sensing wavefront aberrations, with a second ZELDA wavefront sensor (N'Diaye et al., 2016) for improved sensitivity in the high-Strehl regime. Angular Differential Imaging (ADI) will also be employed to reduce the impact of quasi-static speckles. Consequently, the HCAO mode drives the IFS rotator to track the pupil, rather than field tracking as employed in all other modes.

At wavelengths where AO correction is expected to be poor, or when AO cannot be used owing to weather or technical constraints, HARMONI's noAO mode can provide "seeing-limited" performance. The noAO mode utilises a faint ($I < 23$) natural star for slow (~ 0.1 Hz) secondary guiding, eliminating slow drifts of the instrument focal plane and ensuring accurate pointing. This mode is typically expected to be used with the visible grating and the coarsest spaxel scale, as all scales heavily oversample the full width at half maximum (FWHM) of the seeing. 2 × 1 and 4 × 1 binning along the spatial axis can be used to reduce readout times for the CCD detectors, creating effective spaxels of 0.06 × 0.06 arcseconds and 0.06 × 0.12 arcseconds, respectively, that are a better match to the seeing FWHM.

## Instrument description

Figure 2b shows an overview CAD model of the HARMONI instrument. The instrument is ~ 8 m tall, and has a footprint of 5 × 6 m and a total weight of approximately 36 tonnes. The opto-mechanics of the IFS consists of the pre-optics scale changer, the integral field unit (IFU) and four spectrograph units. The IFU rearranges the light from the field into four 500-mm pseudo long slits, which form the input to the four spectrograph units. The IFS opto-mechanics resides in a large cryostat, about 3.26 m in diameter and 4 m tall (a cutaway view is shown in Figure 2a), at a constant operating temperature of 130 K to minimise thermal background. The NIR detectors (eight 4096 × 4096-pixel HAWAII 4RG arrays) are operated at the lower temperature of 40 K. The instrument rotator and cable wrap (IRW) allow the entire cryostat to rotate about a vertical axis to follow field rotation at the ELT's Nasmyth focus. The vertical rotation axis guarantees an invariant gravity vector, improving the instrument's stability by minimising flexure.

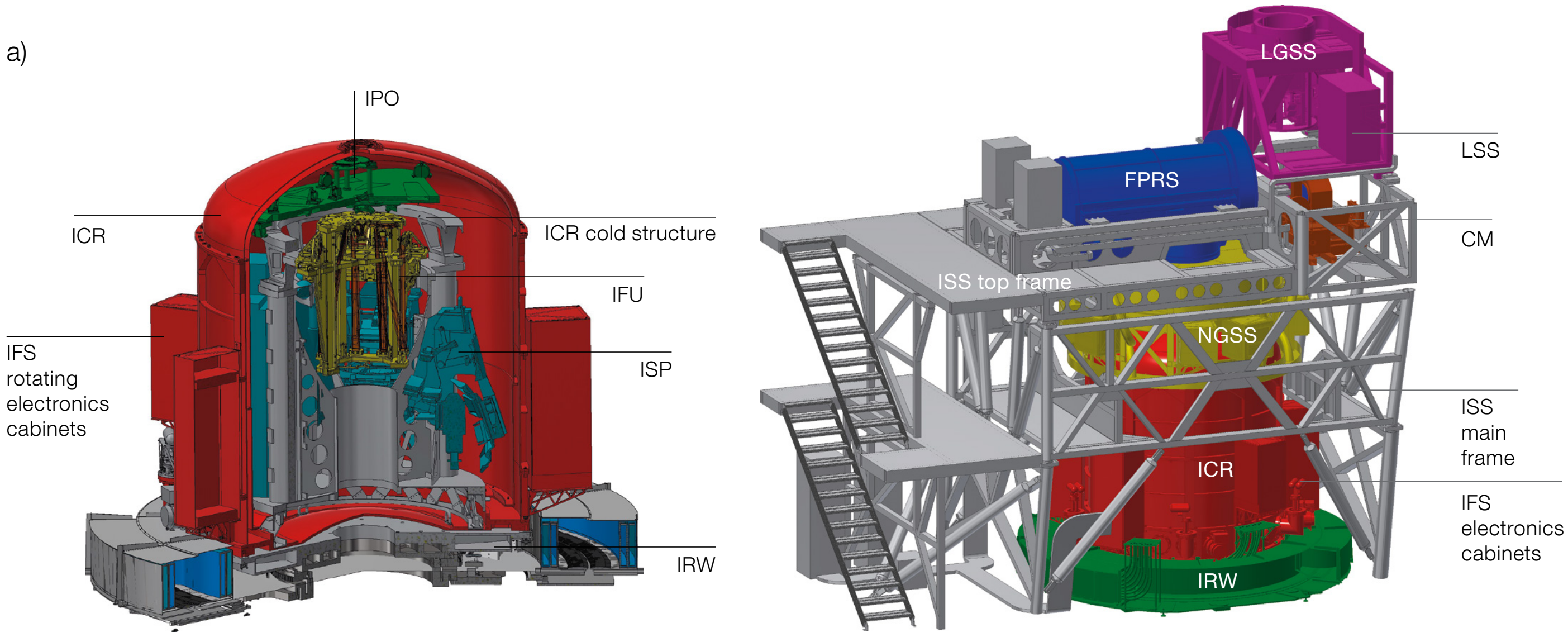

Figure 2. a) Cutaway CAD model of the HARMONI cryostat (ICR), situated on the instrument rotator and cable wrap (IRW). The view shows the main opto-mechanical components of the integral field spectrograph (IFS), namely the IFS pre-optics (IPO), the integral field unit (IFU), and the spectrographs (ISP). b) overall CAD assembly of HARMONI, with the various systems comprising the instrument coloured differently. The LSS is the LGSS Support Structure. Other acronyms are explained in the text.

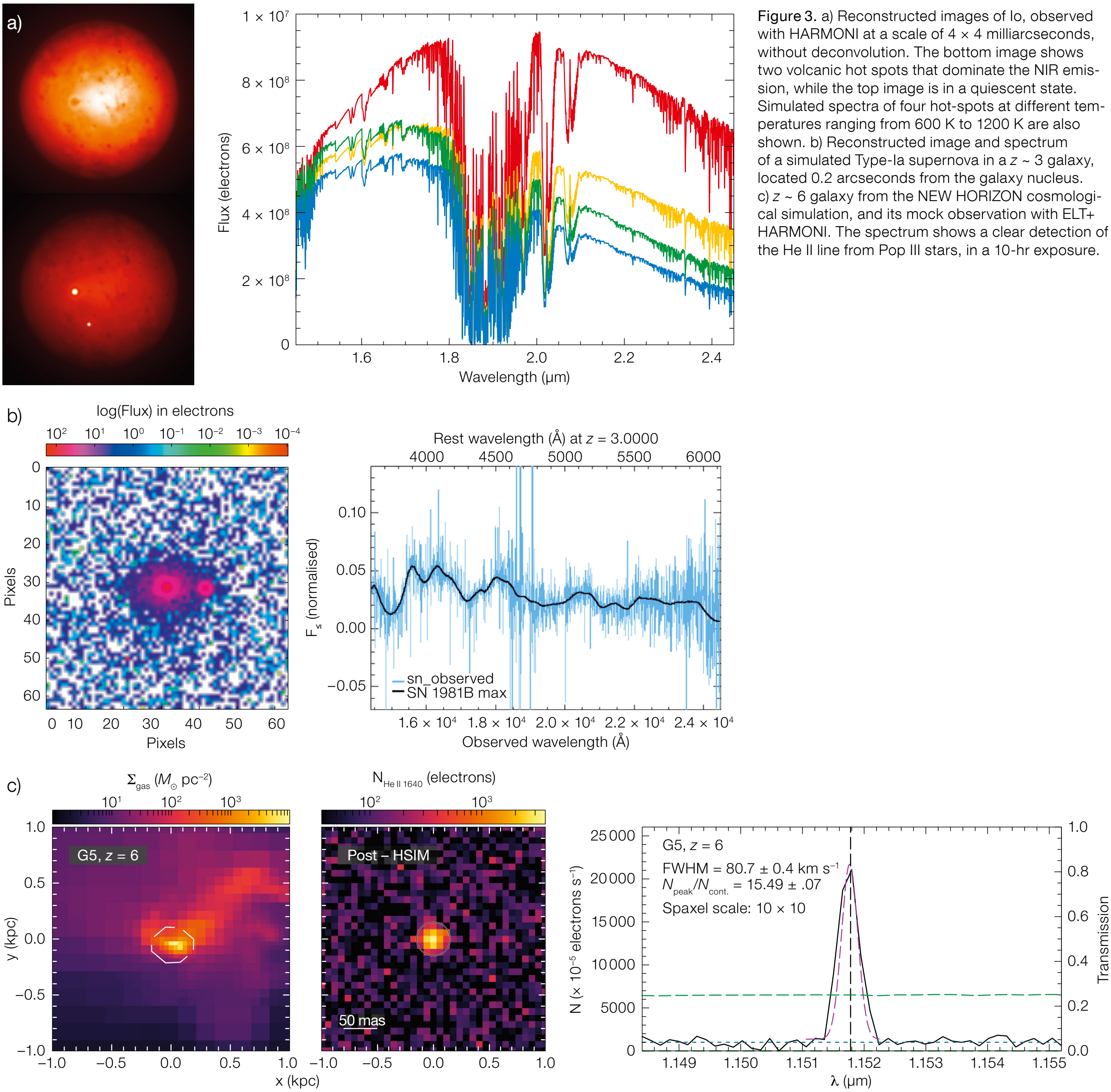

**Figure 3.** a) Reconstructed images of Io, observed with HARMONI at a scale of 4 × 4 milliarcseconds, without deconvolution. The bottom image shows two volcanic hot spots that dominate the NIR emission, while the top image is in a quiescent state. Simulated spectra of four hot-spots at different temperatures ranging from 600 K to 1200 K are also shown. b) Reconstructed image and spectrum of a simulated Type-Ia supernova in a $z \sim 3$ galaxy, located 0.2 arcseconds from the galaxy nucleus. c) $z \sim 6$ galaxy from the NEW HORIZON cosmological simulation, and its mock observation with ELT+HARMONI. The spectrum shows a clear detection of the He II line from Pop III stars, in a 10-hr exposure.

The NGSS is located on top of the IFS cryostat and co-rotates with it. It houses the natural guide star sensors for all four operating modes. As the telescope's back focal distance is insufficient to relay the telescope light directly into the upward-looking cryostat, a focal-plane relay system (FPRS) re-images 2 arcminutes of the telescope focal plane to the top of the cryostat and the NGSS. Both the FPRS and NGSS are maintained in a dry gas environment at a constant temperature of –15 degrees C, reducing thermal background for improved *K*-band sensitivity and minimising thermal drifts.

The LGS System (LGSS) and the Calibration Module (CM) are located just past the instrument slow shutter, close to where telescope light enters the instrument, at a beam height of 6 m above the Nasmyth platform. The first element in the instrument light path is the LGS dichroic, which sends light at 589 nm from the ELT's six LGS to the LGSS. As the LGS asterism is projected from the periphery of the ELT primary mirror (M1), it co-rotates with the telescope pupil, and the LGSS needs its own de-rotator to compensate. The CM can insert light from calibration lamps via fold mirrors into the beam path, mimicking the telescope f-ratio and pupil location. It provides line and continuum sources for all science and technical calibrations. The Instrument Static Structure (ISS) provides a robust mechanical structure and access to all instrument systems.

## Operation and calibration

HARMONI is conceptually simple to operate, as it provides a "point-and-shoot" capability. The user selects one of four operating modes: noAO, SCAO, HCAO or LTAO. In addition, the user must choose a setting that specifies a choice of spaxel scale, grating and, optionally, other user-selectable items (for example, SCAO dichroic, or apodiser) and the instrument is configured accordingly. Accurate pointing is assured by specifying offsets of the science field centre from the natural guide star. As a consequence, the default acquisition sequence does not require an acquisition exposure with the IFS — once the guide star is acquired and all control loops are closed, the first science exposure can commence straight away. Thanks to the unprecedented spatial resolution of the ELT, the accuracy of information needed for guide stars (proper motion, colour, etc.) is much higher than for the Very Large Telescope (VLT). With the faint guide stars which can be used by HARMONI, catalogues may not suffice and pre-imaging of the field might be needed in some cases.

Observing templates will have a similar look and feel to those of other VLT NIR IFS, and will include a variety of sky-subtraction strategies such as "offset to blank sky", "nod-on-IFU" or "stare", together with small jitters to work around bad or hot pixels. Mosaicking will also be supported in the usual way, as will non-sidereal tracking in LTAO and noAO modes (in SCAO and HCAO mode, the only non-sidereal observation possible is when the AO reference "star" is itself non-sidereal). NIR long exposures (typical for spectroscopy of faint targets) will use Sample-Up-The-Ramp (SUTR) readout to minimise readout noise, with every non-destructive readout saved in the archive. AO telemetry data, useful for reconstructing the PSF during the exposure, will also be archived.

Science calibrations needed by the data reduction pipeline, such as arc lamp exposures for wavelength calibration, detector bias and dark frames, flat fields and vertical line and pinhole masks, will be carried out the morning after the observations, as is typical for VLT instruments. ELT instruments are required to be light-tight, so calibrations can happen in parallel for all instruments. With four observing modes, 4 choices of spaxel scale, and 11 grating settings, the number of distinct configurations needing calibration exceeds 100. Consequently, only the configurations used during the night will be calibrated the following morning. Science calibrations and additional monitoring calibrations will be used for "health-checks" (to monitor trends in instrument performance). Efforts will be made to minimise night-time calibrations (telluric or flux standards) wherever possible. Methods that use model-based calibrations instead are being actively investigated by a number of ESO working groups.

## Performance

We have developed a python simulator, HSIM[1], to provide prospective users with the ability to quantitatively assess the efficacy of their proposed observing programme. HSIM (Zieleniewski et al., 2015) is a "cube-in, cube-out" simulator that mimics the effects of atmosphere, telescope, instrument and detector, including the strongly wavelength-dependent, non-axisymmetric AO PSF. The user can analyse the output cube as if it were the output of the instrument pipeline for a real observation, as it incorporates noise from all sources, including shot noise from thermal background and night-sky emission, detector readout noise and dark current. Detector systematics and the impact of sky subtraction can also be included if desired. Through detailed analysis of the output cube, the astronomer can derive uncertainties and confidence levels for the derived physical parameters from the observation, rather than just the signal-to-noise ratio per spaxel (or pixel), thus quantifying the required exposure time or even the feasibility of the planned observation. It also allows the user to develop and test the analysis tools required. The HSIM code is publicly available[2].

HSIM predicts point source sensitivities ($5\sigma$, 5 hr, 2 × 2-spaxel extraction aperture) of $J_{AB} = 25.6$, $H_{AB} = 26.8$, $K_{AB} = 25.9$ in LTAO mode, with SCAO performance of $J_{AB} = 26.2$, $H_{AB} = 27.0$, $K_{AB} = 26.0$ at $R \sim 3000$. The point source sensitivities do not convey the full picture, so we have used HSIM to carry out detailed simulations showcasing a few planned observations with HARMONI. These range from objects in our own Solar System to the most distant galaxies at $z \sim 6$–10.

Jupiter's moon Io is the most volcanically active body in the Solar System. Groussin et al. (in preparation) have simulated ELT observations of Io's hotspots. They show that it is possible to distinguish between sulphurous and ultra-mafic composition of the ejecta by measuring the ejecta's temperature (see Figure 3a) from their NIR spectra, using HARMONI's SCAO mode providing near diffraction-limited spatial resolution.

Bounissou et al. (2018) have shown that HARMONI LTAO can provide direct spectroscopic classification of a supernova in a galaxy at z ~ 3 in a 3-hr observation, up to 2 months past maximum light (see Figure 3b), using the Si II feature (at 400 nm in the rest frame). Confirming type Ia supernovae spectroscopically for a small sub-sample will allow studies of cosmic expansion rates to be pushed to substantially higher redshifts.

We have used the adaptive mesh refinement cosmological simulations from the NEW HORIZON suite (Dubois et al., 2020) to simulate studies of high-$z$ galaxies with HARMONI in a spatially resolved manner. Using cosmological simulations that create galaxies at high spatial resolution commensurate with HARMONI's observational capabilities (~ 100 pc at $z \sim 2$–10) is preferred because the objects have morphologies and kinematic and dynamical properties consistent with the observed ensemble population at high redshifts, and have well understood input physics consistent with known laws and cosmological evolution (Richardson et al., 2020).

Grisdale et al. (2020) have used NEW HORIZON simulations, post-processed using the CLOUDY radiative transfer code (Ferland et al., 2017) to show that HARMONI LTAO could detect the presence of the first stars (Pop III stars) in galaxies at very high redshifts ($z$ = 3–10). The existence of Pop III stars has not been observationally confirmed up to now, although several attempts have been made and some excellent candidates have been identified. Given their primordial composition with no heavy elements, Pop III stars are expected to be substantially more massive than their metal-rich cousins. Consequently, they should burn much hotter, and have a much higher ultraviolet flux, capable of ionising not only hydrogen but also helium in the surrounding gas (H II region). The strength of the He II 164 nm line is thus a good observational diagnostic for the presence of Pop III stars. Despite the large luminosity distance of these very high-redshift star forming regions, the ELT's huge collecting area, coupled with the exquisite spatial resolution provided by HARMONI LTAO, would detect the He II feature with good signal-to-noise from a substantial fraction of the mock galaxies in a 10-hr exposure (Figure 3c). However, to be certain that the line indicates the presence of Pop III stars would require ancillary observations of the H-alpha line from these objects to measure the He II to H-alpha ratio, probably using the James Webb Space Telescope, given the high redshifts involved.

## Acknowledgements

HARMONI work in the UK is supported by the Science and Technology Facilities Council (STFC) at the UK Astronomy Technology Centre (UKATC), Rutherford Appleton Laboratory (RAL), University of Oxford (grants ST/N002717/1 and ST/S001409/1) and Durham University (grant ST/S001360/1), as part of the UK ELT Programme. In France, the HARMONI Project is supported by the CSAA-CNRS/INSU, ONERA, A*MIDEX, LABEX LIO, and Université Grenoble Alpes. The IAC and CAB (CSIC-INTA) acknowledge support from the Spanish MCIU/AEI/FEDER UE (grants AYA2105-68217-P, SEV-2015-0548, AYA2017-85170-R, PID2019-107010GB-100, CSIC-PIE201750E006, and PID2019-105423GA-I00) and from the Comunidad de Madrid (grant 2018-T1/TIC-11035).

The authors would like to acknowledge contributions from Sophie Bounissou (supernova simulations), Olivier Groussin (Io simulations) and Kearn Grisdale (Pop III simulations). We are also grateful to James Carruthers, Neil Campbell, and David Montgomery for CAD views. Miguel Pereira-Santaella is the author of HSIM and we thank him for the sensitivity computations.

## References

Bounissou, S. et al. 2018, MNRAS, 478, 3189
Carlotti, A. et al. 2018, Proc. SPIE, 10702, 107029N
Dubois, Y. et al. 2020, arXiv:2009.10578
Ferland, G. J. et al. 2017, Revista Mexicana de Astronomía y Astrofísica, 53, 385
Grisdale, K. et al. 2021, MNRAS, 501, 5517
N'Diaye, M. et al. 2016, Proc. SPIE, 9909, 99096S
Richardson, M. et al. 2020, MNRAS, 498, 1891
Zieleniewski, S. et al. 2015, MNRAS, 453, 3754

## Links

[1] HSIM simulator: https://harmoni-elt.physics.ox.ac.uk/Hsim.html
[2] HSIM code: https://github.com/HARMONI-ELT/HSIM

## Notes

[a] The full list of HARMONI Consortium members can be found at https://harmoni-elt.physics.ox.ac.uk/consortium.html
[b] Spaxel stands for SPAtial piXEL, to distinguish it from a pixel of the spectrograph detector.

ESO/SPECULOOS Team/E. Jehin

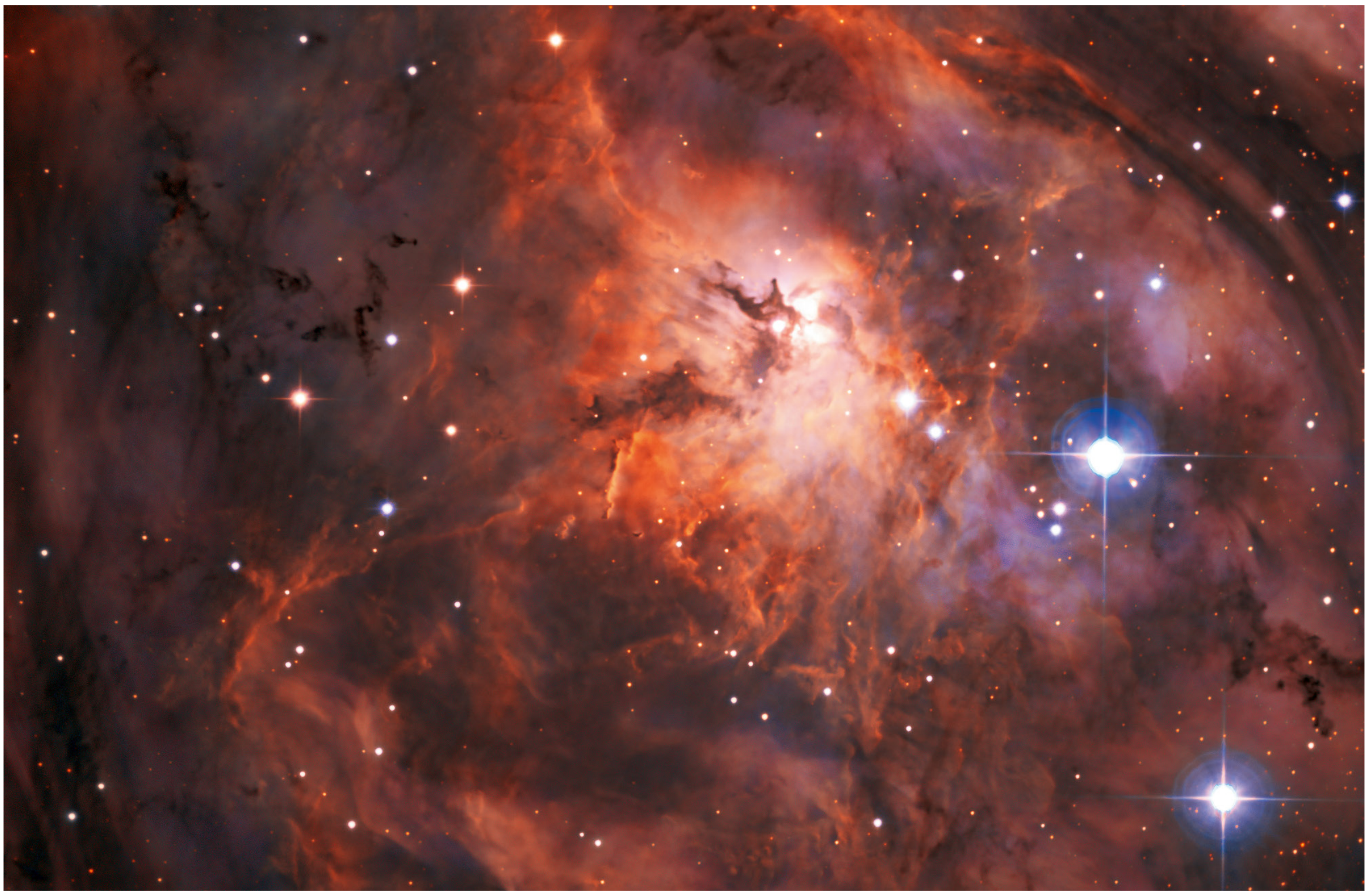

If you had a brand new state-of-the-art telescope facility, what would you look at first? Researchers at the SPECULOOS Southern Observatory — which comprises four small telescopes, each with a 1-metre primary mirror — chose to view the Lagoon Nebula. This magnificent picture is the result, and is one of the SPECULOOS' first ever observations.